\pdfoutput=1
\documentclass[%
reprint,
 amsmath,amssymb,
 aps,
]{revtex4-1}

\usepackage[english]{babel}
\usepackage[utf8x]{inputenc}
\usepackage{mathtools}
\usepackage{amsmath}
\usepackage{multirow}
\usepackage{amsthm}
\usepackage{graphicx}
\usepackage{relsize}
\usepackage{tikz}
\usepackage{verbatim}
\usepackage{pgfplots}
\usepackage{xcolor}
\usepackage{adjustbox}
\pgfplotsset{width=10cm,compat=1.9}
\usepackage[colorinlistoftodos]{todonotes}
\usepackage{hyperref}
\usepackage{chngcntr}
\usepackage{caption} 
\usepackage{subcaption}
\usepackage{tablefootnote}
\usepackage{physics}
\usetikzlibrary{arrows,quotes,angles}
\counterwithin{figure}{section}
\counterwithin{table}{section}

\begin{document}
\newtheorem{definition}{Defintion}
\newtheorem{theorem}{Theorem}
\newtheorem{lemma}{Lemma}
\newtheorem{proposition}{Proposition}

\title{Quantum Mesh Dynamics (QMD)}

\author{Philip Tee}
\email{phil@moogsoft.com}
\altaffiliation[also at ]{Arizona State University, ptee2@asu.edu}
\altaffiliation[and the ]{University of Sussex, p.tee@sussex.ac.uk}

\affiliation{
 Moogsoft Inc and Beyond Center, Arizona State University, \\ 1265 Battery Street, San Francisco, California 94111 USA
}%

\begin{abstract}
A consistent quantum theory of gravity has remained elusive ever since the emergence of General Relativity and Quantum Field Theory.
Attempts to date have not yielded a candidate that is either free from problematic theoretical inconsistencies, falsifiable by experiment, or both.
At the heart of all approaches though the difficult question of what it means for spacetime itself to be quantized, and how that can affect physics, has not been addressed.
In recent years a number of proposals have been made to address the quantum structure of spacetime, and in particular how geometry and locality can emerge as the Universe cools.
Quantum Graphity is perhaps the best known of these, but still does not connect the emerged quantized spacetime to dynamics or gravity.
In this paper we start from a quantized mesh as the pre-geometry of space time and identify that informationally and in a very natural sense, the natural laws of gravity and Newtonian dynamics emerge.
The resultant equations of gravity have a Yukawa term that operates at cosmic scale ($10^{18}$ meters), and we use data from the Spitzer space telescope to investigate experimental agreement of the galactic rotation curves with encouraging results.
We conclude by discussing how this pre-geometry could result in the classical covariant constructs of General Relativity in the low energy continuum limit.
\end{abstract}

\maketitle

\section{Introduction and Background}

Approaches to the reconciliation of General Relativity (GR) with Quantum Field Theory (QFT) have not yielded a consistent and finite theory \cite{Davies1982}.
It is interesting to note that it is fairly well accepted, at least  philosophically, that the concept of a quantized theory of gravity ultimately requires the quantization of space time.
Regardless, contemporary approaches largely ignore that fact and rest solely upon the mathematics of smooth infinitely differentiable manifolds.
Despite increasingly sophisticated mathematical frameworks these approaches usually suffer from an inability to make testable predictions.
For example, String Theory, which essentially argues that the Einstein-Hilbert action of GR \cite{Polchinski2001} is a low energy approximation, utilizes many additional dimensions to yield a finite theory, but as yet has not made contact with experiment.
Further those attempts which directly imply a discrete geometry suffer from a similar absence of experimental support.
Perhaps most well known of these approached is Loop Quantum Gravity \cite{Nicolai2005}.
This approach seeks to divorce the quantization from the background geometry, and implies a spectral and discrete spacetime at the order of the Planck length.
However it again fails to produce testable predictions that admit a satisfactory experimental rigor.

Recently, however, there has been a renewed interest in emergent geometry.
This has its origin in the work of Harland Snyder \cite{Snyder1947}, who proposed a framework to directly consider the implications of discretized space on gravity.
The work was initially dismissed because the existence of a fundamental length implied the existence of preferred observers, which breaks the core principle of Lorentz invariance, and general covariance so vital to the structure of GR.
It has subsequently been proven that no such inconsistency exists \cite{Amelino-Camelia2000,Hossenfelder2012}, and a discrete spacetime can avoid the presence of problematic preferred observers.

If we can accept that spacetime at the order of the Planck Length is fundamentally discrete, this naturally leads to the concept of a mesh or graph, which can be used to base the analysis of gravity as an emergent phenomena.
This approach sometimes termed `Quantum Graphity' \cite{Konopka2008}, provides an intriguing pathway to emergent geometry that  can be shown to naturally require a $3+1$ dimensional universe.
In an elegant fashion Trugenberger combined some recent advances in Network Science to argue how the emergence of a stable graph is a natural phase transition that would have occurred as the universe cooled \cite{Trugenberger2015}, and demonstrated a natural preference for a four dimensional universe.
What these models do not provide, however, is a set of dynamical predictions that can be tested against experiment.

In parallel to these developments, and motivated from the original considerations of Black Hole Thermodynamics \cite{Bekenstein1973,Jacobson1995}, there has been a resurgence in the use of thermodynamics to explain the dynamics of gravity.
In particular, in a series of papers \cite{Verlinde2011,Verlinde2016}, Erik Verlinde has sought to explain the emergence of the classical Newtonian inverse square law as a consequence of the entropy change in a holographic membrane from which our Universe is projected.

In this work, I intend to take a slightly different approach that builds on both the emergent dynamics of Verlinde, and the pre-geometry of Snyder, Konopka {\sl et al} and Trugenberger.
In particular I raise the existence of a fundamental scale and a discrete mesh-like spacetime to a postulate, and ask a very simple and fundamental question, what constraints does this place on the physics occurring on the mesh.
In particular, I consider a spacetime mesh that at each point and edge in the lattice, a Hilbert Space of fermionic spin vectors is overlaid. 
The emergence of space time involves (following Trugenberger) a phase transition to extended local spin alignment and the emergence of a regular graph that exhibits locality (the closest points to any arbitrary point in space are a this points nearest neighbors in the graph, and no connections to distant points exist).
In this space time, a localized anti-alignment involves a state of higher energy than the vacuum, and I identify that with the presence of a localized matter quantum.
Translation of this matter quantum then reduces to the dynamics of edge creation and destruction on this graph.

The inclusion in this model of an emergent gravitational force involves the analysis of the entropy of the underlying spacetime graph.
Graph entropy (for a review see Simonyi \cite{Simonyi1995}), has its origin in the analysis of informational entropy and was introduced by Janos K{\"o}rner \cite{Korner1986} to explain information loss in an imperfect communications channel.
As a consequence it introduced the concept of the structural entropy of a graph that characterizes the lost information in the configuration of a graph when only certain measurable properties of it are accessible. 
In my analysis this results in a net entropy gain when matter particles are adjacent rather than separate and from this an attractive force with the properties of gravity emerge.

Intriguingly, the gravitational force has some subtle difference in the classical limit to GR, in particular at Galactic scale.
It is well known that the dynamics of large scale structures in Cosmology diverge significantly from their predicted behaviors, that resulted in the Modified Newtonian Dynamics (MOND) proposal of Milgrom  \cite{Milgrom1983,Milgrom1983a,Milgrom1983b}.
I apply the latest results of the Spitzer photometry measurements \cite{Lelli2016} to compare the predictive power of this model against a selection of galactic rotation curves, and demonstrate that the QMD model of gravity is at least as efficient as MOND in calculating these rotation curves, but requires less fine tuning of parameters.

\section{The Mesh Model and Dynamics}
\label{sec:mesh}

The {\sl a priori} acceptance of the fundamentally quantum nature of spacetime, and the `background independent' theories of gravity, has a long history.
In our treatment we will focus on attempts to describe spacetime as an emergent order on a spacetime graph.
This approach, often termed Quantum Graphity \cite{Konopka2006,Konopka2008}, has recently seen revised interest in the work of Trugenberger \cite{Trugenberger2015}.
An attractive feature of these theories is the emergence of stable spacetime geometries as the energy and therefore temperature of the Universe drops, which have an intrinsic even dimensionality.
In the case of the Trugenberger model, there is a preferred dimensionality to space of $4$.
Further in Trugenberger's work, it was proven that spacetime can posses topological holes that are stable up to arbitrary temperatures.
The role of time in these models is not specifically stated, and indeed each of the dimensions of the mesh are equivalent.
In this and future work we assume that the dimensions of the mesh are spatial, and time is present in the model as a series of snapshots of the configuration of the mesh.
This has important consequences when we come to consider the continuum limit, as it requires that spacetime acquires an extra degree of freedom to accommodate the notion of time.

Both graph based approaches, however, do not consider dynamics upon the emerged spacetime mesh, or how in the classical or continuum limit a force such as Gravity could emerge.
It is this particular feature we seek to remedy in our treatment.

To begin, we borrow the formalism from Konopka {\sl et al} to create our spacetime, but assume that the ordered mesh has already condensed into a $D=d+1$ dimensional lattice, where $d$ is the spatial dimensionality of the mesh.
We consider a lattice of $N$ nodes, where $N$ is a very large number, representing the quantized positions in space, and we take time to be a represented as a series of configurations of the $1/2N(N-1)$ edges on this graph.
The spacetime graph $G(V,E)$ on this mesh is the paired set of nodes $V$, and the $\frac{1}{2}N(N-1)$ edges $E$ between them.
We adopt standard graph theory notation for a single node $v_i \in V$, and an undirected edge $e_{ij} \in E$ between the vertices $v_i,v_j$, and only consider simple undirected graphs (that is no multiple edges between any pair of nodes, no edges beginning and terminating at the same node, and the assumption thet $e_{ij}$ and $e_{ji}$ are the same edge).
There is an implicit assumption that this mesh nature of spacetime is only evident at very small scales (order of the Plank Length $L_p$), which justifies the very large but finite constraint on $N$ (in \cite{Konopka2008} this number is estimated to be $10^{100}$ to $10^{1000}$).
The requirement of finite N is necessary from graph theoretical considerations.

Over this mesh we associate a Hilbert Space constructed from $N$ fermionic oscillators for the nodes, and $\frac{1}{2}N(N-1)$ such oscillators for the edges.
With this definition the entire Hilbert space is as follows:

\begin{equation}
	\mathcal{H}_{total} = \bigotimes_{1/2N(N-1)} \mathcal{H}_{edge} \bigotimes_N \mathcal{H}_{node}
\end{equation}

The fundamental assumption is that the Hilbert space for the nodes and edges are spanned by two states of a spin 1/2 fermionic oscillator, with which we associate an occupation number of $1$ or $0$, with Fermi-Dirac statistics preventing occupation numbers $ > 1$ and can write:

\begin{equation}
	 \mathcal{H}_{edge/node} = \mbox{span} \{ \ket{0}, \ket{1} \}
\end{equation}

In terms of the spacetime graph, and in classical Quantum Graphity, the focus is upon how a perfect graph where every edge has starting state $\ket{1}$, condenses into a local ordered graph in which only neighbors in the graph have edges, and the nodes have a preferred degree $k_0=2d$.
We shall not reproduce the argument here, but essentially a Hamiltonian is introduced $H=g\sum\limits_a e^{p(k_0-k_a)^2}$, that imposes an energy penalty on nodes in the spacetime graph having a degree different from $k_0$, and also a loop dependent Hamiltonian that penalizes a graph with non-local cycles.

To introduce dynamics, we shall assume that the mesh has condensed at a sufficiently low temperature into an ordered mesh of dimension $D=d+1$.
In this mesh the preferred degree of each vertex $k_0=2d$ represents the ground state of the mesh, and is associated with the matter free vacuum.
We associate departures from $k_0$, in particular values of $k_i<k_0$ as the presence of matter quanta.

At each node $j$ we will associate the state $\ket{v_j,1}$ with the presence of a quantized unit of mass-energy, and $\ket{v_j,0}$ with its absence, and we can write this in terms of the on/off states for each of the edges as follows:

\begin{align}
	\ket{v_j,1} &= \prod\limits_j \ket{e_{ij},0}\\
	\ket{v_j,0} &= \prod\limits_j \ket{e_{ij},1} \mbox{,}
\end{align}
where $j$ ranges over the neighbors of $v_j$ and $\ket{e_{ij},1}, \ket{e_{ij},0}$ are the eigenvectors of the edges in the mesh.
For example in a mesh of $d=2$, the presence of a matter quantum at the node $v_j$, would be characterized by this node having degree $k=0$, which would represent a departure from the ground state by 4 edges.
If $\hat{H}$ is the complete Hamiltonian for the edges, the edges are the eigenvectors of this Hamiltonian, satisfying $\hat{H} \ket{e_{ij},n} = \epsilon \ket{e_{ij},n}$ where $n$ is the occupation number.
For a $d$ dimensional mesh it is therefore natural to associate the quantum of mass-energy $m_q$ contained in this defect, constrained to be $m_q \leq 2d \epsilon/c^2$.

To simplify the argument we define the creation and annihilation operators on a link as follows:

\begin{align}
	a_{ij} \ket{e_{ij},1}&=\ket{e_{ij},0}, a_{ij} \ket{e_{ij},0}=0 \label{eqn:edge_anh} \\
	a^{\dagger}_{ij} \ket{e_{ij},0}&=\ket{e_{ij},1}, a^{\dagger}_{ij} \ket{e_{ij},1}=0 \label{eqn:edge_create} \mbox{,}
\end{align}
and on a node $v_i$ as:
\begin{align}
	a_{i} \ket{1}&=\ket{0}, a_{i} \ket{0}=0 \label{eqn:node_anh}\\
	a^{\dagger}_{i} \ket{0}&=\ket{1}, a^{\dagger}_{i} \ket{1}=0 \label{eqn:node_create} \mbox{,}
\end{align}
where it is understood a single index is a node operator and a double index an operator on the edge represented by the indices.

Finally, we can combine these two ladder operators for the complete operator for the creation and annihilation of matter quantum, localized around the node $v_j$ as:

\begin{align}
	a_{q_i} &= a_i \prod_j a_{ij} \\
	a_{q_i}^{\dagger} &= a_i \prod_j a_{ij}^{\dagger} \mbox{,}
\end{align}
where $j$ ranges over all neighbors of $v_i$.

\begin{figure}
\centering
\begin{adjustbox}{width=\linewidth}
\begin{tikzpicture}
	\node [black] at (0,1) {\textbullet};
	\draw (0,1) node [above] {$x-1$};
	\draw (0,1) node [left] {$y+1$};
	\node [black] at (1,1) {\textbullet};
	\draw (1,1) node [above] {$x$};
	\node [black] at (2,1) {\textbullet};
	\draw (2,1) node [above] {$x+1$};
	\node [black] at (3,1) {\textbullet};
	\draw (3,1) node [above] {$x+2$};
	
	\node [black] at (0,0) {\textbullet};
	\draw (0,0) node [left] {$y$};
	\node [black] at (1,0) {\textbullet};
	\node [black] at (2,0) {\textbullet};
	\node [black] at (3,0) {\textbullet};
	
	\node [black] at (0,-1) {\textbullet};
	\draw (0,-1) node [left] {$y-1$};
	\node [black] at (1,-1) {\textbullet};
	\node [black] at (2,-1) {\textbullet};
	\node [black] at (3,-1) {\textbullet};
	
	\draw (1,0) node [below=0.25cm] {$(x,y)$};
	\draw (1,-1) node [below=0.25cm, right] {$t=t_0$};
	
	\draw [-] (0,0) -- (0,1);
	\draw [-] (0,0) -- (0,-1);
	\draw [-] (0,1) -- (1,1);
	\draw [-] (0,-1) -- (1,-1);
	\draw [-] (1,-1) -- (2,-1);
	\draw [-] (2,-1) -- (3,-1);
	\draw [-] (1,1) -- (2,1);
	\draw [-] (2,1) -- (3,1);
	\draw [-] (2,0) -- (2,1);
	\draw [-] (2,0) -- (2,-1);
	\draw [-] (2,0) -- (3,0);
	\draw [-] (3,0) -- (3,1);
	\draw [-] (3,0) -- (3,-1);

	\draw [->] (4,0) -- (5,0) node [above left=0.3cm] {$\hat{T_x}$};
	
	\draw [->] (4,-2) -- (5,-2) node [right] {$x$};
	\draw [->] (4,-2) -- (4,-1) node [above] {$y$};
	
	\node [black] at (6,1) {\textbullet};
	\draw (6,1) node [above] {$x-1$};
	\draw (6,1) node [left] {$y+1$};
	\node [black] at (7,1) {\textbullet};
	\draw (7,1) node [above] {$x$};
	\node [black] at (8,1) {\textbullet};
	\draw (8,1) node [above] {$x+1$};
	\node [black] at (9,1) {\textbullet};
	\draw (9,1) node [above] {$x+2$};
	
	\node [black] at (6,0) {\textbullet};
	\draw (6,0) node [left] {$y$};
	\node [black] at (7,0) {\textbullet};
	\node [black] at (8,0) {\textbullet};
	\node [black] at (9,0) {\textbullet};
	
	\node [black] at (6,-1) {\textbullet};
	\draw (6,-1) node [left] {$y-1$};
	\node [black] at (7,-1) {\textbullet};
	\node [black] at (8,-1) {\textbullet};
	\node [black] at (9,-1) {\textbullet};
	
	\draw [green,-] (6,0) -- (7,0);
	\draw [green,-] (7,0) -- (7,1);
	\draw [green,-] (7,0) -- (7,-1);
	\draw [-] (6,0) -- (6,1);
	\draw [-] (6,0) -- (6,-1);
	\draw [-] (6,1) -- (7,1);
	\draw [-] (6,-1) -- (7,-1);
	\draw [-] (7,-1) -- (8,-1);
	\draw [-] (8,-1) -- (9,-1);
	\draw [-] (7,1) -- (8,1);
	\draw [-] (8,1) -- (9,1);
	\draw [red,-] (8,0) -- (8,1);
	\draw [red,-] (8,0) -- (8,-1);
	\draw [red,-] (8,0) -- (9,0);
	\draw [-] (9,0) -- (9,1);
	\draw [-] (9,0) -- (9,-1);

	\draw (8,0) node [below=0.25cm, fill=white] {$(x+1,y)$};
	\draw (7,-1) node [below=0.25cm, right] {$t=t_0+1$};
\end{tikzpicture}
\end{adjustbox}
\caption{Diagrammatic effect of a translation $\hat{T_x}$, when $D=2+1$. Edges that are annihilated are depicted in red, created in green.}
\label{fig:translation}
\end{figure}
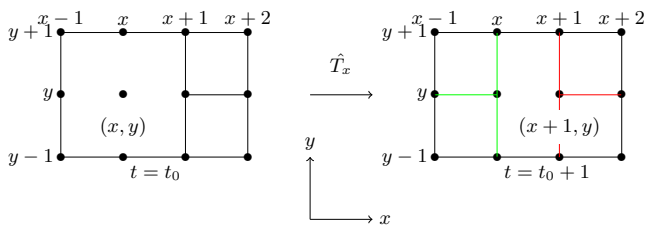

In Figure \ref{fig:translation} we depict the operation of a translation operator $\hat{T}_x$ in the $x$ direction on a mesh of dimension $2+1$.
To model the effect of the translation we need to associate directionality into the structure of the mesh.
As noted before, the graph is technically finite, but is assumed to be very large.
If we overlay upon the graph the notion of independent directions, we can do so by associating with each node $v_j$ a coordinate $x_i=(x,y)$.
For ease of notation, we will assume that each node has a unique coordinate label, and further assume in our example drawn in Figure \ref{fig:translation}, that the node one mesh distance in the positive $x$ direction, for example, is uniquely labelled with coordinate $(x+1,y)$.
To model a translation occurring in one time step, the node $v_i$ located at position $(x,y)$, must gain edges, and the node $v_j$ at the position $(x+1,y)$ must lose some.
This is indicated in the diagram by coloring the annihilated edges in red and the created edges in green in the diagram.
The two states of the mesh are separated by one time step, the duration of which we can adjust to preserve the normal constancy of the speed of light.
If $d$ denotes the spatial dimensionality of the mesh (i.e. $D=d+1$), the translation requires the annihilation of $2d-1$ edges and creation of $2d-1$. 
To simplify the notation, we introduce the following notation for edge adjacency creation/annihilation operators, for a translation in the $x$ direction as:

\begin{align}
	\overleftarrow{a}_{(x,y)}(\mathbf{x}) &= a_{(x,y),(x-1,y)} a_{(x,y),(x,y-1)} a_{(x,y),(x,y+1)} \\
	\overleftarrow{a}_{(x,y)}^{\dagger}(\mathbf{x}) &= a_{(x,y),(x-1,y)}^{\dagger} a_{(x,y),(x,y-1)}^{\dagger}  a_{(x,y),(x,y+1)}^{\dagger}  \\
	\overrightarrow{a}_{(x,y)}(\mathbf{x}) &= a_{(x,y),(x+1,y)} a_{(x,y),(x,y-1)} a_{(x,y),(x,y+1)} \\
	\overrightarrow{a}_{(x,y)}^{\dagger} (\mathbf{x}) &= a_{(x,y),(x+1,y)}^{\dagger}  a_{(x,y),(x,y-1)}^{\dagger}  a_{(x,y),(x,y+1)}^{\dagger} 
\end{align}

Using this more compact notation we can write the translation operator at point $(x,y)$ in terms of the operators in Equations \eqref{eqn:edge_anh},\eqref{eqn:edge_create},\eqref{eqn:node_anh},\eqref{eqn:node_create} as follows:

\begin{equation}
	\hat{T}_x^{(x,y)} = : a^{\dagger}_{x+1} \overleftarrow{a}_{(x,y)}^{\dagger}(\mathbf{x}) a_x \overrightarrow{a}_{(x+1,y)}(\mathbf{x}) :
\end{equation}

From here we can begin to estimate the relationship between energy, force and motion of this quantum on the mesh.
We begin by asserting the minimal distance on the mesh, the spatial `length' of the edges in the graph to be identified with the Planck Length $L_p=1.6 \times 10^{-35}m$ \cite{Garay1995}.
From the original Hamiltonian for the mesh, every departure of the node from its preferred degree carries an energy penalty.
We can associate this energy penalty with the occupation number of an edge, and denote the energy input required to annihilate an edges as $\epsilon$.
In this way when an edge is created the energy of the system decreases by $\epsilon$, and when an edge is annihilated it correspondingly increases by $\epsilon$.
In this model the notion of the vacuum is the minimum energy state of the mesh, with no holes and therefore total matter quantum count of zero.
Above this vacuum energy in the mesh will result in the evolution of defects or matter quanta.

A localized matter quantum at rest in the mesh, is therefore stable and will remain localized in a $d$ dimensional volume $(2L_p)^d$.
To model a matter quantum in uniform motion, conceptually one considers a potentially infinite series of translations, with a velocity of $\mathbf{v}$, being to be comprised of $|\mathbf{v}|/L_p$ such translations in unit time.
During a translation there is a net energy input required of $(2d-1)\epsilon$ to annihilate the forward links (equating to a departure from the preferred degree $k_0$), which is offset by the same amount of energy released when the back links are created.

That is the action of $\overrightarrow{a}_{(x+1,y)}(\mathbf{x})$ consumes the $(2d-1)\epsilon$ units of energy liberated by the action of $\overleftarrow{a}_{(x,y)}^{\dagger}(\mathbf{x})$ on the mesh for motion in the $x$ direction.
In this uniform motion the quantum, as a closed system, has no input or output of energy to continue its motion, at least when averaged over time.
In this way we arrive at Newton's first law of motion, that a body in uniform motion will continue so in the absence of external forces.

For a quantum moving at velocity in excess of one mesh step per unit time interval, and respecting the normal ordering of the edge annihilation/creation operators, the motion requires an $(2d-1)\epsilon|\mathbf{v}|/L_p$ investment in energy to annihilate forward edges before that energy is recovered from edge creation.
To preserve causality this energy cannot be transferred instantaneously, and so in some way constitutes an energy related to the motion of the quantum that we can associate with the kinetic energy of the quantum.

To consider acceleration, let us assume that the quantum is in uniform motion at a velocity of $v_0$ requiring no external input of energy, averaged over a time period large compared to $L_p/v_0$. 
We now imagine a force $F$ being applied to the mass-energy quantum for a time $\delta t$.
As the particle is traveling at velocity $v_0$, this will result in a total input energy $\Delta E=F v_0 t$.
This energy will permit the lattice to annihilate an additional $\frac{\Delta E}{(2d-1)\epsilon}$ forward edges in $\delta t$, by applying additional $\overrightarrow{a}_{(x,y)}(\mathbf{x})$ operators.
At velocity $v_0$, the mass energy quantum is applying $\overrightarrow{a}_{(x+1,y)}(\mathbf{x})$ at $v_0 /L_p$ lattice positions in a unit time interval.
As we noted previously before the net energy requirement a constant velocity is zero, but if energy is invested this can step change up the number of $\overrightarrow{a}_{(x,y)}(\mathbf{x})$ operators in the unit time interval resulting in an increase in velocity.
We denote this increase in lattice position moves in the time interval $\delta t$ as $\delta q$, which is related to a change in velocity as $\delta v = L_p \delta q$.
Additionally for a localizable quantum (up to the uncertainty principle) the mass-energy $m_q$ must also be bounded above such that $m_q < 2d\epsilon/c^2$.
Appealing to the conservation of energy we now have:

\begin{align}
	Fv_0\delta t &= \frac{1}{2} m_q\Bigg \{ (v_0+L_p \delta q )^2 -v_0^2 \Bigg \} \\
	Fv_0\delta t &= \frac{1}{2} m_q ( 2 v_0 L_p \delta q + O(\delta q^2) )  \mbox{, ignoring $\delta q^2$,  } \\
	F		   &= m_q\frac{ \delta v } {\delta t } \mbox{, or familiarly}\\ \label{eqn:qmd_accel}
	F		   &= m_q a \mbox{.}
\end{align}

The defect propagation model on the mesh has delivered the familiar $1^{st}$ and $2^{nd}$ laws of motion, with translation being stable in the absence of external forces and forces causing the translational velocity to accelerate in proportion to the number of mass energy quanta being accelerated.

There is however one further and very subtle consequence of Equation \eqref{eqn:qmd_accel}, namely that there is a minimum amount of acceleration possible in this system.
Until the force applied can generate an additional $(2d-1)\epsilon$ of energy over $\delta t$, {\sl no acceleration is possible}. 
This introduces a minimum acceleration into the system, and in principle at low enough impulse, not sufficient to generate a quantum of translational energy $\epsilon$, bodies should remain in uniform motion.
This is in complete contrast to the normal rules of Newtonian dynamics, which permits an arbitrarily small acceleration in response to an arbitrarily small applied force.
The concept of a Modified Newtonian Dynamics (MOND) has been advanced to explain the anomalous behavior of galaxies at very low acceleration that result in  the flat rotation curves of galaxies \cite{Milgrom1983,Milgrom1983a,Milgrom1983b}.
In MOND,  the rotational velocity of stars at the extreme edge of galaxies appears to be faster than necessary to balance the gravitational attraction from the observed mass of the galaxy.
QMD however offers another explanation, as it is entirely natural that the dynamical behavior on the mesh at very low values of acceleration (and by implication energy) would be different than at normal scales.
In MOND there is a characteristic acceleration scale $a_0$, at which the dynamics is proposed to depart significantly from Newtonian. 
It is conceivable that the minimum acceleration scale emerging from the dynamics of the mesh may be approached as an asymptote dynamically at very small accelerations.
\section{Emergence of Gravity}
\label{sec:gravity}

\subsection{From Quantum Mesh to Gravitation}

In the previous section we established that a modification to Quantum Graphity, allows us to treat the fermionic spin states on the edges of the mesh as a representation of a confined matter quantum.
Specifically we modeled a hole in the mesh as representing a position in spacetime containing a matter quantum.
From this simple assumption we were able to derive simple analogs of the laws of Newtonian motion.
In this section, we plan to extend the program to ask if this structure can be used to identify any emergent forces that behave like gravity.

Our starting point is the informational content of the mesh, by treating it as a finite (though very large) graph.
Any graph has an associated structural entropy that can be defined upon it \cite{Korner1986,Simonyi1995,Dehmer2011,Anand2011,Tee2017}, although there are many variants of how that can be done.
In particular, it is possible, at least approximately, to reduce the definition of graph entropy to a sum over the constituent node's entropy, that becomes exact when the graph is highly regular (\cite{Dehmer2011,Tee2017, Tee2018}.
This closest form of vertex entropy to the behavior of structural entropy \cite{Tee2018}, is defined in terms of the degree $k$ of each vertex, and the total number of edges in the graph $|E|$.
As the graph is extremely large, this number of edges can be assumed to be a constant for spacetime, which for later analysis we can safely ignore, as we are only concerned with entropy changes as matter quanta coalesce. 
For dimension $d$ it is defined as:

\begin{equation}\label{eqn:vertex}
	S_v(k) =-\frac{k}{|E|} \log_2 \frac{k}{|E|} = \frac{2d}{|E|} \log_2 \frac{|E|}{2d} \mbox{.}
\end{equation}

It is already well known that there is a close correspondence between informational entropy, traditional thermodynamic entropy and GR.
The concept has been used to  explain  black hole thermodynamics \cite{Bekenstein1973}, the field equations of General Relativity as state equations \cite{Jacobson1995}, and also the emergence of gravity \cite{Verlinde2016,Hossenfelder2017}.
The recent work by Verlinde, is intriguing but requires the acceptance of a holographic universe, and a correspondence between matter and information directly, a concept for which there is no experimental evidence.

Our starting point is the emerged quantized fermionic mesh, which requires that the universe is spatially $4$-dimensional (with the assumption that there is an additional time-like dimension representing the sequence of configurations of the mesh).
In addition, if we can explain a gravity like force emerging from the maximization of the entropy of this mesh it will immediately posses several important properties.
Firstly, and unique amongst the known forces, gravity always acts in an attractive manner between masses.
The $2^{nd}$ law of thermodynamics likewise has no reverse gear, so in a closed system (such as the Universe) systems tend towards the overall maximization of entropy.
Secondly, gravity is a linear function of the amount of `matter' that is gravitating, and in a similar way informational entropy is additive so the more quanta that are present the more entropy and potential force from maximization is possible.
In a dynamic process the $2^{nd}$ law of thermodynamics will maximize entropy, and essentially that is our entry point to analyze the emergence of an attractive force between matter quanta on the mesh.
In simple terms our hypothesis is that the entropy maximization force and the attractive force of gravity are equivalent.

\begin{figure}
\centering
\begin{adjustbox}{width=\linewidth}
\begin{tikzpicture}
	\node [black] at (0,1) {\textbullet};
	\node [black] at (1,1) {\textbullet};
	\node [black] at (2,1) {\textbullet};
	\node [black] at (3,1) {\textbullet};
	\node [black] at (0,0) {\textbullet};
	\node [black] at (1,0) {\textbullet};
	\node [black] at (2,0) {\textbullet};
	\node [black] at (3,0) {\textbullet};
	\node [black] at (0,-1) {\textbullet};
	\node [black] at (1,-1) {\textbullet};
	\node [black] at (2,-1) {\textbullet};
	\node [black] at (3,-1) {\textbullet};
		
	\draw [-] (0,0) -- (0,1);
	\draw [-] (0,0) -- (0,-1);
	\draw [-] (0,1) -- (1,1);
	\draw [-] (0,-1) -- (1,-1);
	\draw [-] (1,-1) -- (2,-1);
	\draw [-] (2,-1) -- (3,-1);
	\draw [-] (1,1) -- (2,1);
	\draw [-] (2,1) -- (3,1);
	\draw [-] (3,0) -- (3,1);
	\draw [-] (3,0) -- (3,-1);
	
	\node [black] at (5,1) {\textbullet};
	\node [black] at (6,1) {\textbullet};
	\node [black] at (7,1) {\textbullet};
	\node [black] at (8,1) {\textbullet};
	\node [black] at (9,1) {\textbullet};
	\node [black] at (10,1) {\textbullet};	
	\node [black] at (5,0) {\textbullet};
	\node [black] at (6,0) {\textbullet};
	\node [black] at (7,0) {\textbullet};
	\node [black] at (8,0) {\textbullet};
	\node [black] at (9,0) {\textbullet};
	\node [black] at (10,0) {\textbullet};
	\node [black] at (5,-1) {\textbullet};
	\node [black] at (6,-1) {\textbullet};
	\node [black] at (7,-1) {\textbullet};
	\node [black] at (8,-1) {\textbullet};
	\node [black] at (9,-1) {\textbullet};
	\node [black] at (10,-1) {\textbullet};
	
	\draw [-] (5,1) -- (6,1);
	\draw [-] (6,1) -- (7,1);
	\draw [-] (7,1) -- (8,1);
	\draw [-] (8,1) -- (9,1);
	\draw [-] (9,1) -- (10,1);
	
	\draw [-] (5,-1) -- (6,-1);
	\draw [-] (6,-1) -- (7,-1);
	\draw [-] (7,-1) -- (8,-1);
	\draw [-] (8,-1) -- (9,-1);
	\draw [-] (9,-1) -- (10,-1);
	
	\draw [-] (5,0) -- (5,1);
	\draw [-] (5,0) -- (5,-1);
	\draw [red,-] (7,0) -- (7,1);
	\draw [red,-] (7,0) -- (7,-1);
	\draw [red,-] (7,0) -- (8,0);
	\draw [-] (8,0) -- (8,1);
	\draw [-] (8,0) -- (8,-1);
	\draw [-] (10,0) -- (10,1);
	\draw [-] (10,0) -- (10,-1);
\end{tikzpicture}
\end{adjustbox}
\caption{Mesh configuration when two matter quanta are in contact, and separated. On the right hand side the edges marked red are net edge losses. The presence of a matter quantum is signified by an isolate degree zero vertex.}
\label{fig:vertex}
\end{figure}
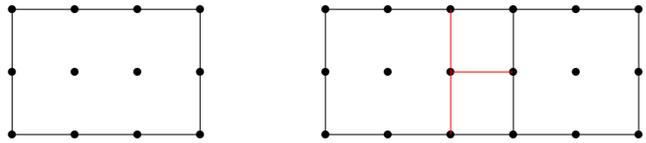

If we consider Figure \ref{fig:vertex} we can see how the configuration of the mesh changes as two matter quanta move from being separated to being in contact.
The presence of a matter quantum is characterized by a zero degree node contained in the mesh defect.
On the left hand side the two quanta have coalesced, whereas on the right hand side they are separated by one mesh step of $L_p$.
The key point is that in \eqref{eqn:vertex} a node of degree $2d-1$ has a slightly lower entropy than one of degree $2d$, that is the presence of a matter quantum has an entropy lowering effect on the entropy of the whole mesh.
This is caused by the $2d$ number of degree $2d-1$ nodes.
The constrained nodes with $k=0$, do not contribute to the entropy of the mesh as they are unchanged in number, and in any case the value of \eqref{eqn:vertex} as $k \rightarrow 0$ is zero.
When the matter quanta are coalesced, this number of lower degree nodes decreases by $d$, and therefore relative to the mesh the entropy drop is lower for two quanta to be in contact rather than separate, and the two nodes being contained in the same defect is therefore an entropy gain overall for the mesh.
 This is consequently favored as an equilibrium configuration.
 It is also interesting to note in passing that the number of lower degree nodes is a function of the size of the $d-1$ boundary of the defect, rather than its volume, which is a result consistent with the well known laws of  black hole entropy.

This increase in entropy permits the definition of an entropic force, as it is well known that a system in thermal equilibrium can exert a restoring force when a mechanical process would decrease the systems entropy.
The most used example of these are polymer chains \cite{Verlinde2011}, and osmosis, but it is also possible to use this to explain other phenomena.
Indeed amongst the first of such attempts to explain physical phenomenon in this emergent was Brownian motion in the breakthrough work by Neumann \cite{Neumann2005} and then Roos \cite{Roos2013}.
In a similar way this author previously treated the dynamical evolution of graphs and the emergence of scale freedom using similar arguments \cite{Tee2017b}.
In all cases the process that changes the system entropy at temperature $T$, generates a force $F$, in the direction of increase of entropy according to:

\begin{equation}
	F=T \Delta S
\end{equation}

The entropy maximization assumes that all matter quanta are perfectly packed, but there is of course a maximum density of matter in the mesh and would not correspond to the more normal gravitational sources.
In essence, as a mass-energy quantum moves from a region of higher to lower density the probability of two matter quanta being in direct contact will decrease linearly with the density of matter.
As such the effective drop in mesh entropy as a quantum separates will also depend upon the density of the gravitating matter, a fact that will become important when we test the resultant entropic force against observational data.
For now, we will denote the maximum matter density as $\rho_m$, which corresponds to a region of space where every mesh cell is occupied.
Correspondingly the vertex entropy increase as quanta are separated $\delta S_g$,  will need to be adjusted for the density $\rho$ of the gravitating body $\delta S_g \propto \delta S_{m} \times \frac{\rho}{\rho_m}$.
For example, the average density of the region between our Sun and the nearest star can be computed to be approximately $1.6 \times 10^{-15} kgm^{-3}$, by distributing the mass evenly through that volume.
This can be converted to a thermodynamic entropy drop $S_q$, by multiplying by the Boltzmann constant $k_B$, and we should note that for $d=3$ the dimensions of this entropy change would be $JT^{-1}L^{-3}$.
The precise form of the entropy relationship is not important for calculation purposes as it will only appear as a multiplicative constant in any estimations, and in any case \eqref{eqn:vertex} is dependent upon the total number of edges in the quantum mesh, a finite but potentially large number.
For the purposes of our later analysis we can assume that as the size of the total number of edges is effectively constant, the overall drop in entropy will be proportional to the Boltzmann constant in magnitude as matter quanta are separated. 

We initiate our analysis by considering two matter quanta separated by a distance $r$.
Because of the inherent spherical symmetry of this system from the perspective of either quanta, we can consider one particle fixed at the origin, and the other at this distance $r$.
We wish to ask the question how does the entropy $S(r)$ of the mesh change as the value of this separation $r$ changes.
Once we have an expression for that, we can establish the force acting on the quanta, in the direction $\hat{\mathbf{r}}$ using similar arguments to the work of Verlinde and Roos \cite{Verlinde2011,Verlinde2016,Roos2013} as:

\begin{equation}\label{eqn:entropic}
	F_{ \hat{\mathbf{r}} } = T \frac{ \partial S(r) } { \partial r } \mbox{.}
\end{equation}

We know that when the two quanta are in contact we know that the entropy of mesh increases by $\delta S_g$, but to recover a distance dependence upon the entropy of the mesh, we need to relate this value to the overall entropy of the system.
Given that the Universe has ambient thermal energy, we approach that question by asking what is the probability is of a random thermal fluctuation causing a translation that  brings the two quanta together and thereby increasing the entropy of the system.
If we denote the probability of this transition as $P(\hat{T}_r)$, then the entropy of the mesh $S(M)$ on average changes by $\delta S(M)=P(\hat{T}_r) \times \delta S_g$).
The quantum mesh is assumed to be in a heat bath of temperature $T$, which can provide energy to support random fluctuations in the mesh.
We have already established that to generate a translation of length $r$, we need to provide an energy input of $r(2d-1)\epsilon/L_p$, sufficient to annihilate directionally the edge quanta on the path to $r$ from the origin.
In the subsequent analysis we fix the value of $d=4$, and so the energy input is $7\epsilon r/L_p$.
We can decompose the probability of the translation $P(\hat{T}_r)$ into two factors.
The first is simply $L_p^2/4\pi r^2$, being the probability of randomly choosing a direction for the translation that causes the two matter quanta to become adjacent.
The second is the probability of a thermal fluctuation producing enough energy to generate the translation.

This second term we can compute by treating every translation of one mesh position as a discrete energy level in a system obeying Fermi-Dirac statistics, and using standard results to compute the probability of the matter quantum translating spontaneously a distance equivalent to that energy level.
For economy of notation we define two constants:
\begin{equation} \label{eqn:a}
	a=\frac{7 \epsilon  }{ L_p k_B T} \mbox {,}
\end{equation}
and for convenience later:
\begin{equation}\label{eqn:z}
	z=\frac{7 \epsilon  }{ L_p k_B} \mbox {.}
\end{equation}
This allows us to write for this probability $P_{\mbox{therm}}(\hat{T}_r)$ as:
\begin{equation}
	P_{\mbox{therm}}(\hat{T}_r) = \frac{1}{Z} \frac{1}{1+\exp(ar)} \mbox{,}
\end{equation}
where $Z$ is the partition function, defined as:

\begin{equation}
	Z = \sum\limits_{r/L_p=0}^{r/L_p=\infty} \frac{1}{1+\exp(ar)}
\end{equation}

We can compute the value of the partition function by approximating the sum as an integration over all space, which permits us to write $ \frac{1}{1+\exp(ar)} \approx e^{-ar}$.
We therefore compute:

\begin{equation*}
	Z=\int\limits_0^{2\pi} \int\limits_0^\pi  \int\limits_0^\infty r^2 e^{-ar} \sin \theta  dr d\theta d\phi \mbox{.}
\end{equation*}
This is easily integrated, and yields for our partition function:

\begin{equation}\label{eqn:part_fn}
	Z=\frac{ 8 \pi  } {a^3} \mbox{.}
\end{equation}

We note that for a body of mass $M$, composed of $M/m_q$ quanta, the total entropy drop will be simply factored by $M/m_q$.
We can now assemble our expression for the entropy drop associated with two matter quanta being separated by a distance $r$ at an ambient mesh temperature of $T$:

\begin{equation}\label{eqn:ent_r}
	S(r)=\frac{S_q a^3 L_p^2 M}{32 \pi^2 m_q} \times \frac{1}{(1+e^{ar})r^2}
\end{equation}

We can now apply \eqref{eqn:entropic} to obtain the entropic force bringing the two quanta together, but before we do so it is possible to further simplify our expression.
The exponential term in \eqref{eqn:ent_r} must for relatively small (i.e. at the scale of the solar system) be effectively a constant, otherwise the entropic force would be inconsistent with Newtonian, and indeed relativistic forms of gravitational attraction. This implies that $ar \ll 1$, and to all intents and purposes, we can approximate the denominator as $(1+e^{ar})^{-1} \approx 1-e^{ar}$.
Once we apply substitute and differentiate we obtain for values of $ar \ll 1$:

\begin{equation}\label{eqn:force_full}
	F_{ \hat{\mathbf{r}} } = - \frac{ S_q L_p^2 T M }{32 \pi^2 m_q} \times \Bigg ( \frac{ a^4 e^{ar}}{r^2} +  \frac{ 2a^3 (1-e^{ar})}{r^3}\Bigg)\mbox{. }
\end{equation}

The second term, will introduce an additional restoring force, which will rapidly drop off with distance, and would result in the same mass appearing to gravitate more strongly than the expected Newtonian effect.
Of course this is the restoring force exerted by a mass of $M$ located at the origin on a single matter quanta at distance $r$.
To compare this with traditional Newtonian forces, we need to consider a second composite mass at distance $r$.
If we denote the active gravitational mass $M_1$, and the passive mass $M_2$, we have:

\begin{equation}\label{eqn:force}
	F_{ \hat{\mathbf{r}} } = - \frac{ S_q L_p^2 T M_1 M_2 }{32 \pi^2 m_q^2} \times \Bigg ( \frac{ a^4 e^{ar}}{r^2} +  \frac{ 2a^3 (1-e^{ar})}{r^3}\Bigg)\mbox{. }
\end{equation}

This formula is none other than the familiar Newtonian inverse square law, with a Yukawa potential correction and, intriguingly, a temperature dependent coupling constant, and a second more complex $r^{-3}$ term that will be negligible for $ar \ll 1$.
To perform a gross error check on the model, we note that the dimensions of the gravitational coupling constant $G$ are $M^{-1}L^3T^{-2}$.
Our expression for $G(T)$ (ignoring the second term in \eqref{eqn:force} as it is dimensionally identical to the first, and can be ignored for small $r$) is:

\begin{equation}
	G(T)=\frac{ S_q L_p^2 a^4 T }{32 \pi^2 m_q^2} \mbox{.}
\end{equation}

As $TS_q$ has dimensions of $JL^{-3}$, and $a^4$ dimensions of $L^{-4}$, and $m_q$ has dimensions of $ML^{-3}$,with energy having dimensions of $ML^2T^{-2}$, we have:

\begin{align*}
	\mbox{ Dimensions: } G(T)&=\frac{ ML^2T^{-2} L^{-3} L^2 L^{-4} }{ M^2 L^{-6}} \\
						 &= M^{-1}L^3T^{-2} \mbox{ , }
\end{align*}
as required, indicating that our expression for $G(T)$ is dimensionally consistent.

To further investigate the temperature dependence of the coupling constant, we need to expand out $a^4$ in terms of $z$  as it contains $T$ from Equation \ref{eqn:a} and \ref{eqn:z}.
When we do this  the term simplifies to:

\begin{equation*}
	G(T) = \frac{ S_q L_p^2 z^4}{32 \pi^2 m_q^2 T^3} \mbox{ .}
\end{equation*}

This, however, ignores the temperature dependence inherent in the Yukawa term of Equation \ref{eqn:force}.
To demonstrate the effect, let us consider at a fixed distance $R$ from a mass $M$, the effective coupling constant would have the following temperature dependence:

\begin{equation}\label{eqn:G}
	G(T) = \frac{ S_q L_p^2 z^4}{32 \pi^2 m_q^2 } \times \frac{ \exp( \frac{- z R }{T}) } { T^3}  \mbox{ .}
\end{equation}

This variance in the coupling constant is complex, and has two asymptotic regions where $G(T) \rightarrow 0$, where $T \rightarrow \infty$ and $T \rightarrow 0$ as depicted in Figure \ref{fig:coupling}.
Both of those temperature regimes have significant consequences on the evolution of the Universe, as temperature decreased from an initial very high value to the CMB temperature of $2.73K$. 
In particular, it would indicate that initially gravity was very weak, and strengthened as the Universe cooled, potentially opening the door to a mechanism to explain the arrest of inflationary cosmic models.

\subsection{Implications for Large Scale Structure}

The form of Equation \eqref{eqn:force_full} has some interesting implications for the large scale structure of the Universe. 
In particular it introduces two new and distinct modifications to the gravitational attraction that modify the effect of the force of gravity both at very small scales and very large ones.

At the small scale, the derivative term which has an $r^{-3}$ distance dependency looks at first a little unusual.
However, if you consider General Relativity, the Schwarzchild metric introduces precisely such a term.
Following Rindler \cite{Rindler2001}, the metric produces an the following approximate gravitational force law per unit mass:

\begin{align}
	F_{ \hat{\mathbf{r}} }&=- \frac{G}{r^2}\Bigg(1-\frac{2G}{c^2r} \Bigg)^{-1/2}\\
					& \approx - G\Bigg (  \frac{1}{r^2} + \frac{G}{c^2r^3} + \dots \Bigg )
\end{align}

At the large scale, Equation \ref{eqn:force_full}  also introduces an exponential correction with a characteristic length of $1/a$.
As it is anticipated that $a$ is very small, this characteristic length could be at least of the order of galactic scale, and approximately unity for smaller distance that operate for at the solar system range.
This opens up the possibility of using constants determined using local observations terrestrially, to test the form of Equation \ref{eqn:force_full}  using the more recent observations of large scale galactic dynamics, such as the Spitzer SPARC dataset of 175 galaxies \cite{Lelli2016}.

We begin by using Equation \ref{eqn:G} to calculate the approximate size of the scale factor $z$, by considering a stellar mass, such as our Sun, located at $r=0$ and use the measured form of $G$ terrestrially. 
For the background temperature of the mesh we use the temperature of the CMB of $2.73^\circ K$.

Once we have established the temperature of the gravitating body, to make use of Equation \ref{eqn:G}, we  have to also calculate the value of $S_q$, the amount of entropy that is lost when the mass-energy network is altered by thermal excitation of the mesh translating the mass-energy quantum to a position $r$ distant from the gravitating mass at position $r=0$. 
We can use Equation \ref{eqn:vertex} to estimate this entropy drop, assuming that the mass-energy quantum is transported from a fully packed mesh to free space, and then factor this by the degree of packing of the mesh at the gravitating source.
When we come to apply this to galactic dynamics, this will become critical and we will need to factor in the relative density of matter across a galaxy.

As noted earlier, we make the assumption that the entropy drop involved in separating two matter quanta is proportional to the Boltzmann constant factored for the density of gravitating material relative to a fully packed mesh.
To estimate the critical density of a fully packed mesh involves a numerical iteration to solve for a value of $z$, which in turn yields a value for $\epsilon$, as from \eqref{eqn:z} $\epsilon=k_B L_p z/7$.
Once we have $\epsilon$, in turn we can estimate $m_q$, by using the estimate that $m_q < 2d\epsilon/c^2$, and then $\rho_m$ as $m_q/L_p^3$.
At a given temperature, we can therefore continuously vary the value of $z$ until we arrive at a value for $G=6.67 \times 10^{-11} m^3kg^{-1}s^{-2}$ the known local value, when we do so we arrive at $z=4.24 \times 10^{-21} Km^{-1}$.
We summarize the other relevant values arising from this calculation in Table \ref{tab:solvedG}.

\begin{table}
\centering
\begin{tabular}{|c|c|c|c|}
\hline
&&&\\[-11pt]
$z$ & $\epsilon$ & $m_q$ & Scale Factor $1/a$  \\[2pt] \hline
&&&\\[-9pt]
$4.24\times10^{-21}$ & $1.78\times10^{-79}$ & $1.19 \times10^{-95}$ & $6.43\times10^{20}$  \\[2pt] \hline
\end{tabular}
\caption{Estimate of critical parameters to reproduce $G=6.67\times10^{-11}m^3kg^{-1}s^{-2}$ at Earth.}
\label{tab:solvedG}
\end{table}

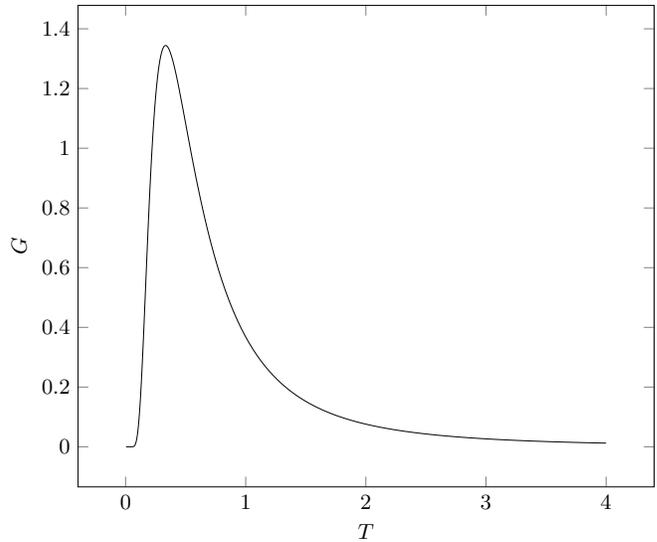
\begin{figure}
\centering
\begin{adjustbox}{width=\linewidth}
\begin{tikzpicture}
\begin{axis}[xlabel={$T$}, ylabel={$G$}]
	\addplot[domain=0:4, samples=1000] {exp(-1/x)/(x^3)};
\end{axis}
\end{tikzpicture}
\end{adjustbox}
\caption{Variation of $G(T)$ as $T$ varies for Fixed Distance from a Gravitating Mass.}
\label{fig:coupling}
\end{figure}

This set of estimates is intriguing because it introduces a natural scale factor into the effects of the quantum mesh on the force of gravity of $6.43\times10^{20} m$, or approximately $20$ kPc, which is of the order that will produce significant distances at galactic and intergalactic scales.
The temperature dependence of $G$ is also an interesting feature of the model, and we plot the form of that relationship at a fixed distance from a gravitating body in Figure \ref{fig:coupling}.
The principle feature is that at zero temperature the coupling is zero, and also at high temperatures it asymptotically approaches zero.
For a `hot big bang' model this would imply that initially gravity is very weak, but that as the Universe cools it passes through a maximum.
Intriguingly that would admit a model of Universe formation where the effect of gravitation would vary as the Universe cooled.
Initially the force would be very weak, permitting rapid expansion, and then would peal as the Universe cools, potentially quite rapidly.
This may serve as an explanation for how inflation was an early Universe phenomenon that halted as the Universe cooled.

Given the natural scale of the QMD gravitational force operates at cosmological scale, it is natural to test the operation of this force against available datasets, such as the SPARC dataset.
We do so in the next section.

\section{Large Scale Tests of QMD}
\label{sec:experimental}

\subsection{Methods and Data}
It has been well known since the early observations of Vera Rubin and the work of  Fritz Zwicky \cite{Zwicky1937} that the observed mass of galaxies does not explain their rotational dynamics.
In particular the observed mass is insufficient to account for the total rotational acceleration, which led ultimately to the competing explanations of the so called `Dark Matter' and MOND \cite{Milgrom1983,Milgrom1983a,Milgrom1983b}.
Other attempts, such as emergent gravity \cite{Lelli2017}, have tried to tackle the requirement to input non-universal parameters into MOND and also areas where MOND does less well, but there is still no clear resolution to the Dark Matter debate.
Accordingly it is natural to use this particular problem to test the validity of the QMD expression for gravitational attraction, as the scale factors computed in the previous section would operate at the scale of a typical galaxy.

The source data used for comparison is drawn from the SPARC collaboration, built upon observations using the Spitzer space telescope \cite{Lelli2016}.
The data covers 175 galaxies of many different morphologies and compositions, and is the most accurate measurements of rotation curves and surface luminosity available.
From this dataset it is possible to estimate the mass distribution of Baryonic material from the surface luminosity $\Upsilon$, and Mass-to-Light ratios (see for example \cite{McGaugh2011}) of the galaxies, and from there numerically solve for the rotation curves of them.
In common with all studies of rotation curves from observational data \cite{Lelli2016,Lelli2017,Bhattacharjee2014,McGaugh2008,McGaugh2014}, we will make the normal assumptions of circular orbits, and uniform smooth distribution of the Baryonic matter.

It was previously noted in Section \ref{sec:gravity}, that the entropy drop is dependent upon the matter density difference between the active and passive gravitational masses, and because of the temperature dependence of the gravitational coupling, we should also take into account how the temperature of the galaxy systems also vary with density.
In our computational model, we construct a ratio $\Phi=\Upsilon(r)/\Upsilon_{m}$, where $\Upsilon_m$ is the maximum surface luminosity of the galaxy, and $\Upsilon(r)$ is the surface luminosity at distance $r$ from the center of the galaxy.
This can be used as a proxy for the density differences in the distribution of stellar material.
The connection between density and temperature of the mesh is surely complex, but we will make a gross assumption of linearity which could be improved upon in future analysis.
Accordingly we use this ratio to scale both the effective temperature $T_e$  and the effective entropy $S_{e}$, which we parameterize as follows:

\begin{align}\label{eqn:parameters}
	S_{e}&=\frac{S_g}{\Phi} \\
	T_{e}&=(\alpha + \beta \Phi)T_s \mbox{.}
\end{align}

We can use the \eqref{eqn:parameters} along with Equations \eqref{eqn:ent_r}, \eqref{eqn:entropic} to numerically integrate to solve for the acceleration and therefore rotational velocities of the galaxies studied in the SPARC survey.
The data in SPARC consists of bolometric surface density measurements and rotational velocity obtained from $H_1$,$H_\alpha$ studies.
To convert the luminosity measurements to observable Baryonic mass, the Galaxy supplement data from the same source \cite{Lelli2016}, has a table of conversion factors.
Once the velocity dispersion curves are computed, an overall numerical optimization is undertaken to minimize the total variance between the computed curve and the measured one, by introducing a variable factor to $\Upsilon(r)$, which is a constant for all points in the galaxy.
This does not affect the form, shape or correlation of the computed curve and amounts to a global adjustment of the mass to light ratio for the galaxy.
This same factor is applied to a computation using the normal Newtonian expression for gravitational attraction.

The code written to perform all of the analysis was implemented in Matlab, and is available on request from the author.

\subsection{Comparison of Computed Rotation Curves}

\begin{figure*}[t]
	\centering
	\begin{subfigure}[t]{0.45\textwidth}
		\centering
		\includegraphics[scale=0.4]{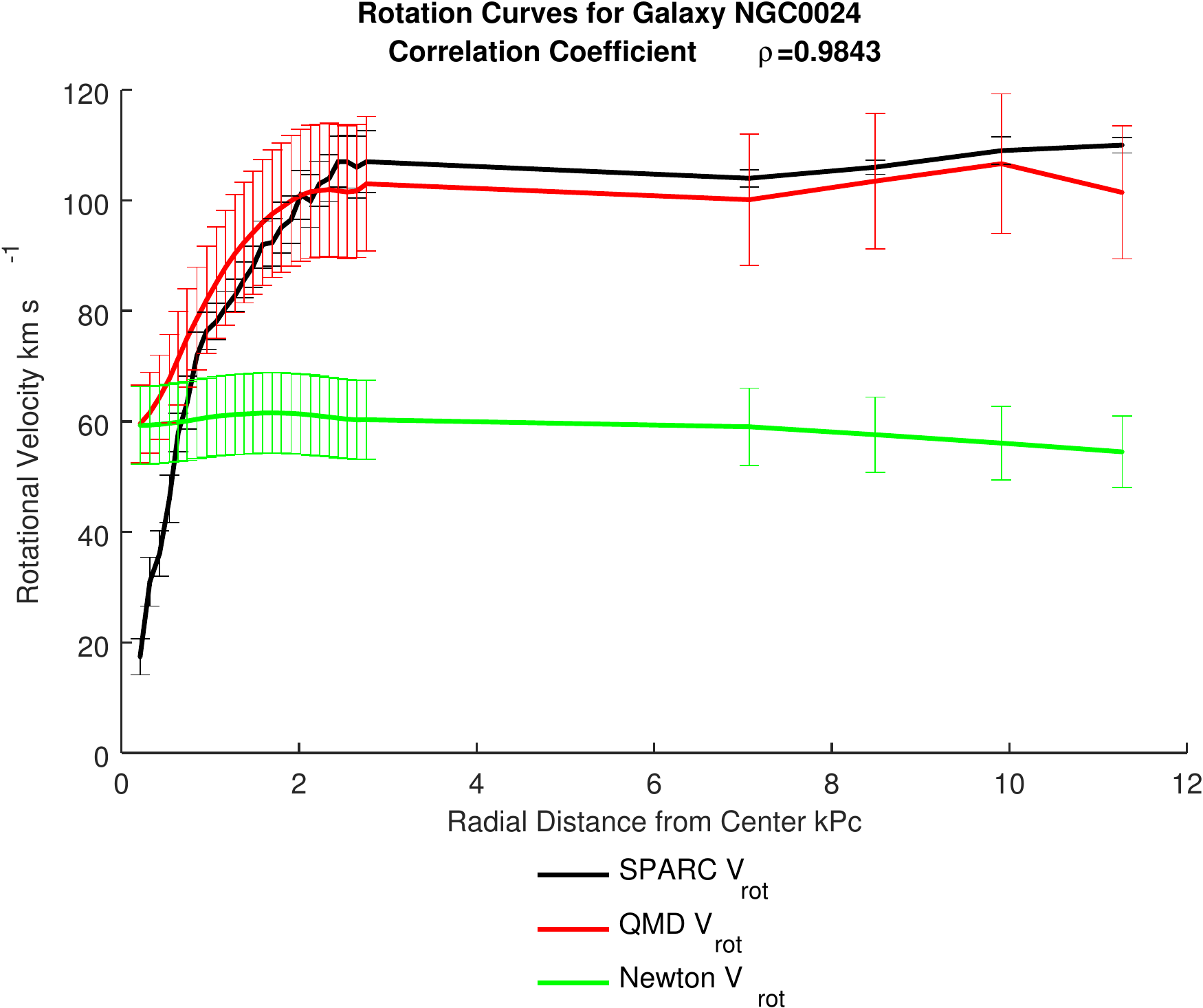}
		\caption{Galaxy NGC 0024}
		\label{fig:NGC0024}
	\end{subfigure}%
	~ 
	\begin{subfigure}[t]{0.45\textwidth}
		\centering
		\includegraphics[scale=0.4]{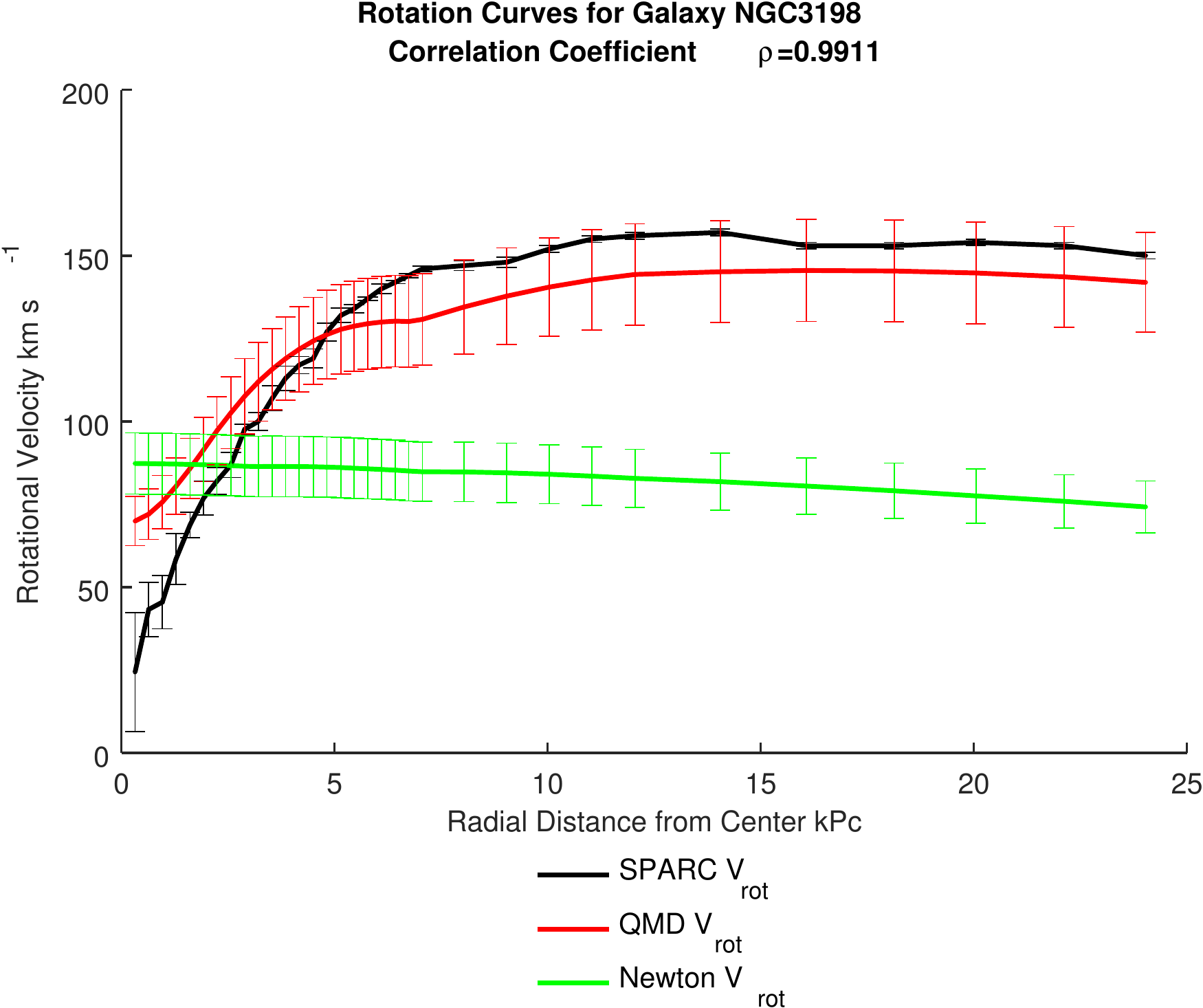}
		\caption{Galaxy NGC 3198}
		\label{fig:NGC3198}
	\end{subfigure}
	~ 
	\begin{subfigure}[t]{0.45\textwidth}
		\centering
		\includegraphics[scale=0.4]{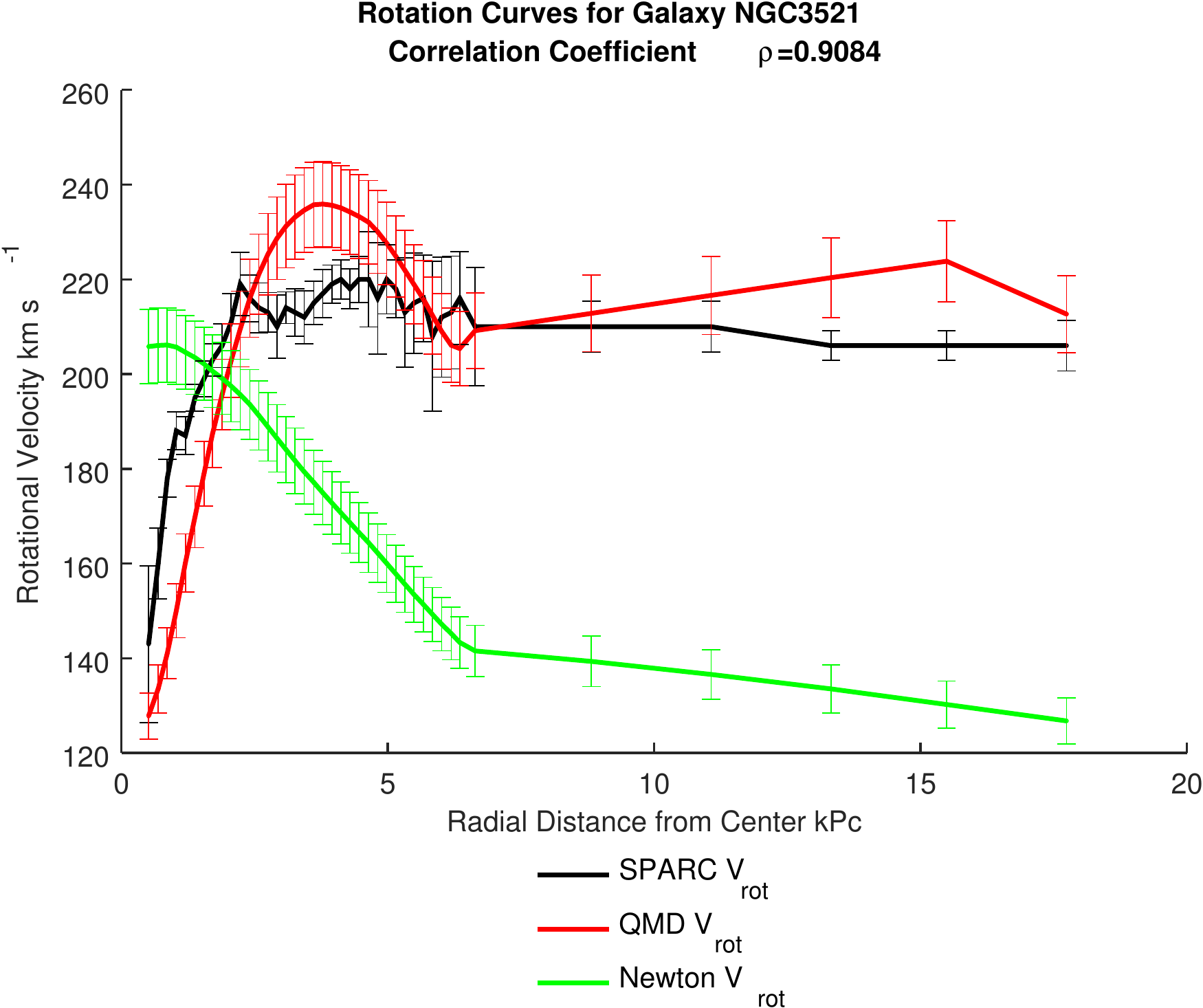}
		\caption{Galaxy NGC 3521}
		\label{fig:NGC3521}
	\end{subfigure}
	~ 
	\begin{subfigure}[t]{0.45\textwidth}
		\centering
		\includegraphics[scale=0.4]{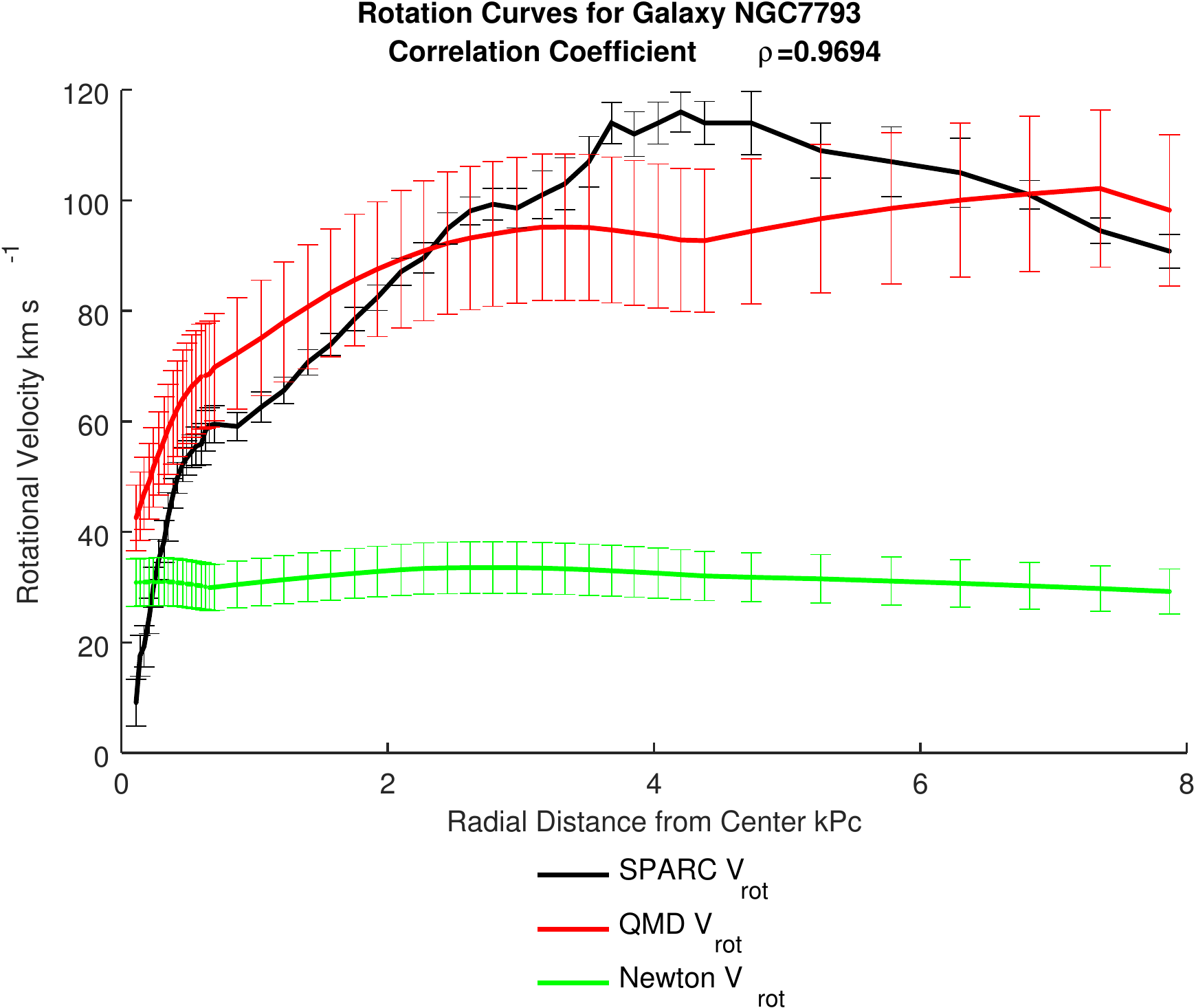}
		\caption{Galaxy NGC 7793}
		\label{fig:NGC7793}
	\end{subfigure}
	~ 
	\begin{subfigure}[t]{0.45\textwidth}
		\centering
		\includegraphics[scale=0.4]{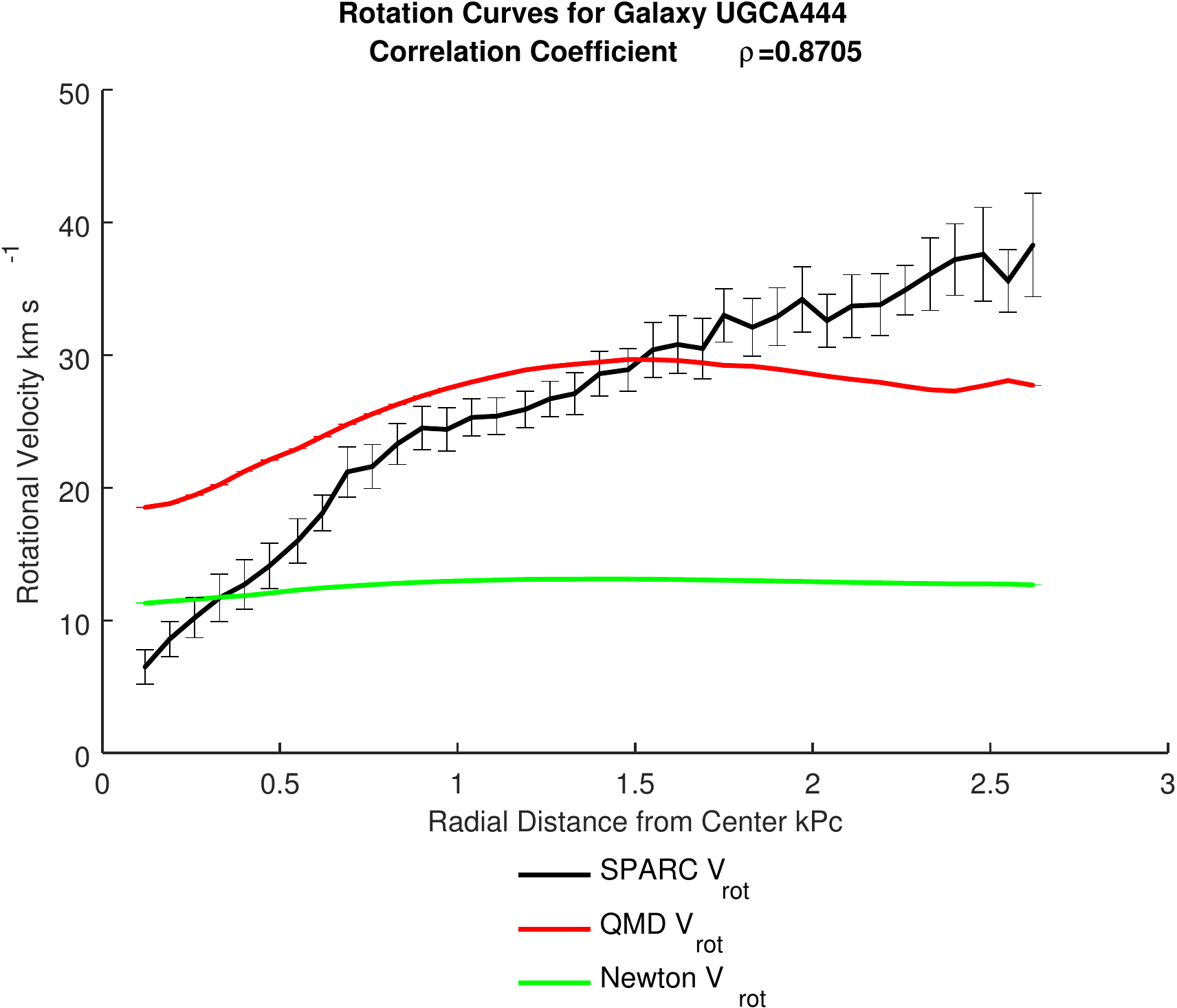}
		\caption{Galaxy UGC A444}
		\label{fig:UGCA444}
	\end{subfigure}
	~ 
	\begin{subfigure}[t]{0.45\textwidth}
		\centering
		\includegraphics[scale=0.4]{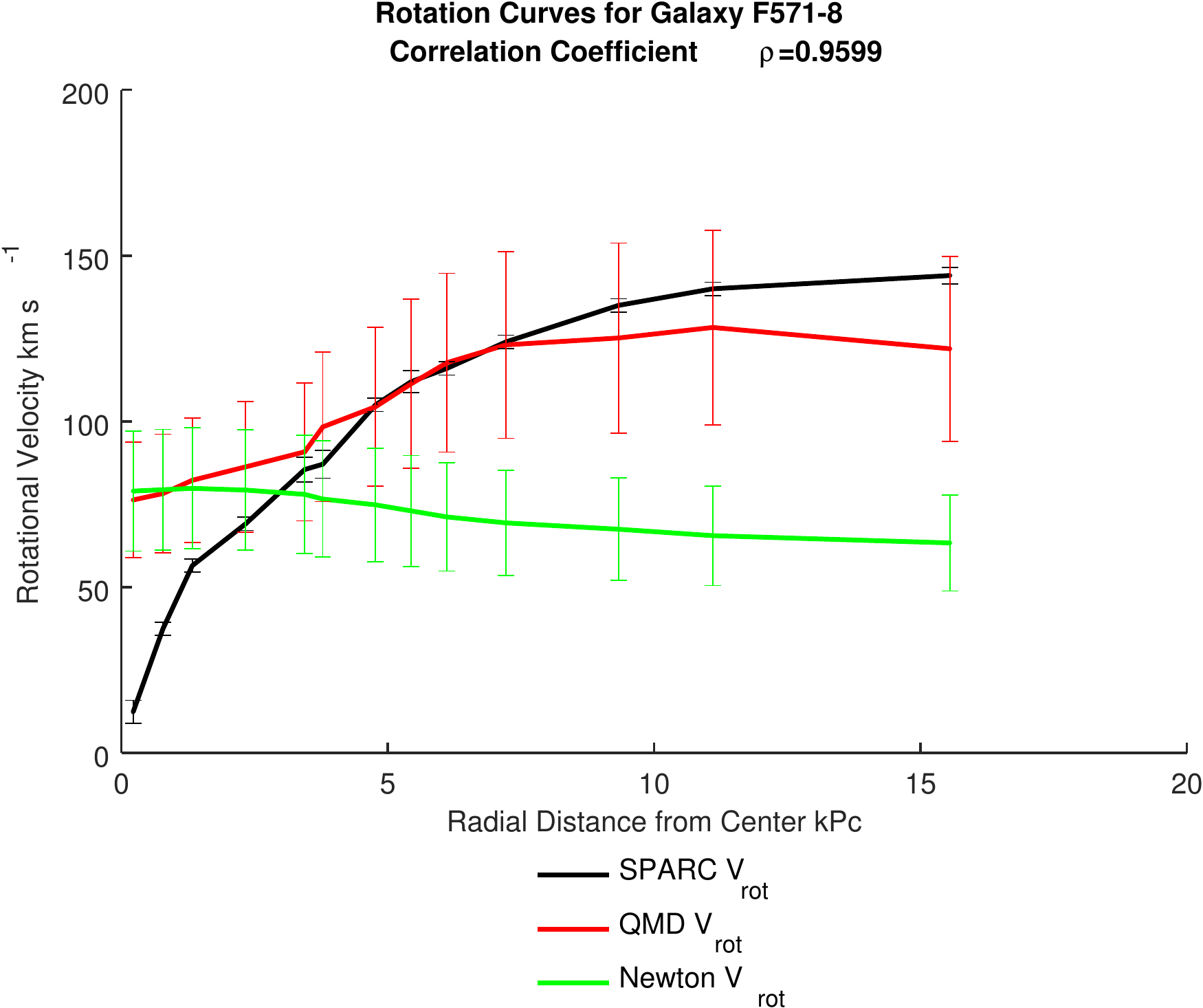}
		\caption{Galaxy F571-8}
		\label{fig:F571-8}
	\end{subfigure}
	\caption{A selection of computed rotation curves using \eqref{eqn:force_full} and standard Newtonian $r^{-2}$ force law.}
	\label{fig:rotation}
\end{figure*}

In Figures \ref{fig:NGC0024},\ref{fig:NGC3198},\ref{fig:NGC3521},\ref{fig:NGC7793},\ref{fig:UGCA444}, \ref{fig:F571-8} we present a selection of computed and measured rotation curves.
This covers a range of galaxy morphology and types.
NGC7793 and NGC24 are both spiral galaxies in the Sculptor constellation, NGC3521 is a barred spiral galaxy in the constellation of Leo,  NGC3198 a barred spiral in Ursa Major, UGCA444 is an irregular galaxy in our local group, and F571-8 an edge-on elliptical galaxy.
All of the depicted galaxies are low surface brightness galaxies typically studied to debate the effectiveness of MOND over dark matter.

In solving numerically the best fit using \eqref{eqn:parameters} we applied the same constant multiplicative factor to the Newtonian mass density as the QMD calculation.
The measured rotation curve is the thick black line, QMD is the red line and Newtonian is the green plot.
Error bars are computed by taking the measurement error in the surface luminosity and carrying through the relevant equation to depict the translated error in the computation.
Our best fit for the data involved a value of $\alpha=1.0$, $\beta=1.0$, a modest variance in temperature between the centers of the galaxies above the $2.73^{\circ}K$ used for the ambient temperature of the mesh.

\begin{figure*}[t]
	\centering
	\begin{subfigure}[t]{0.45\textwidth}
		\centering
		\includegraphics[scale=0.4]{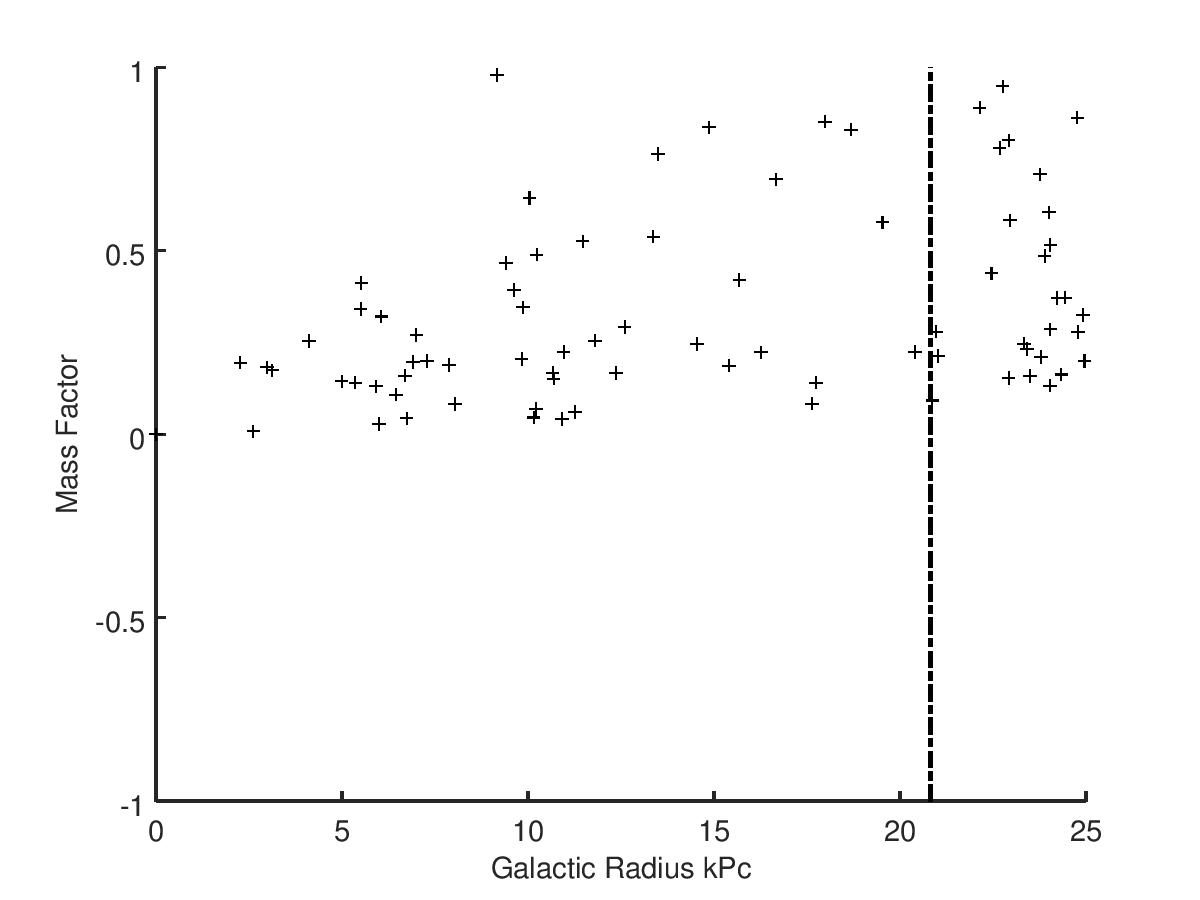}
		\caption{Computed Mass Factor versus Galaxy Radius}
		\label{fig:MF}
	\end{subfigure}%
	~ 
	\begin{subfigure}[t]{0.45\textwidth}
		\centering
		\includegraphics[scale=0.4]{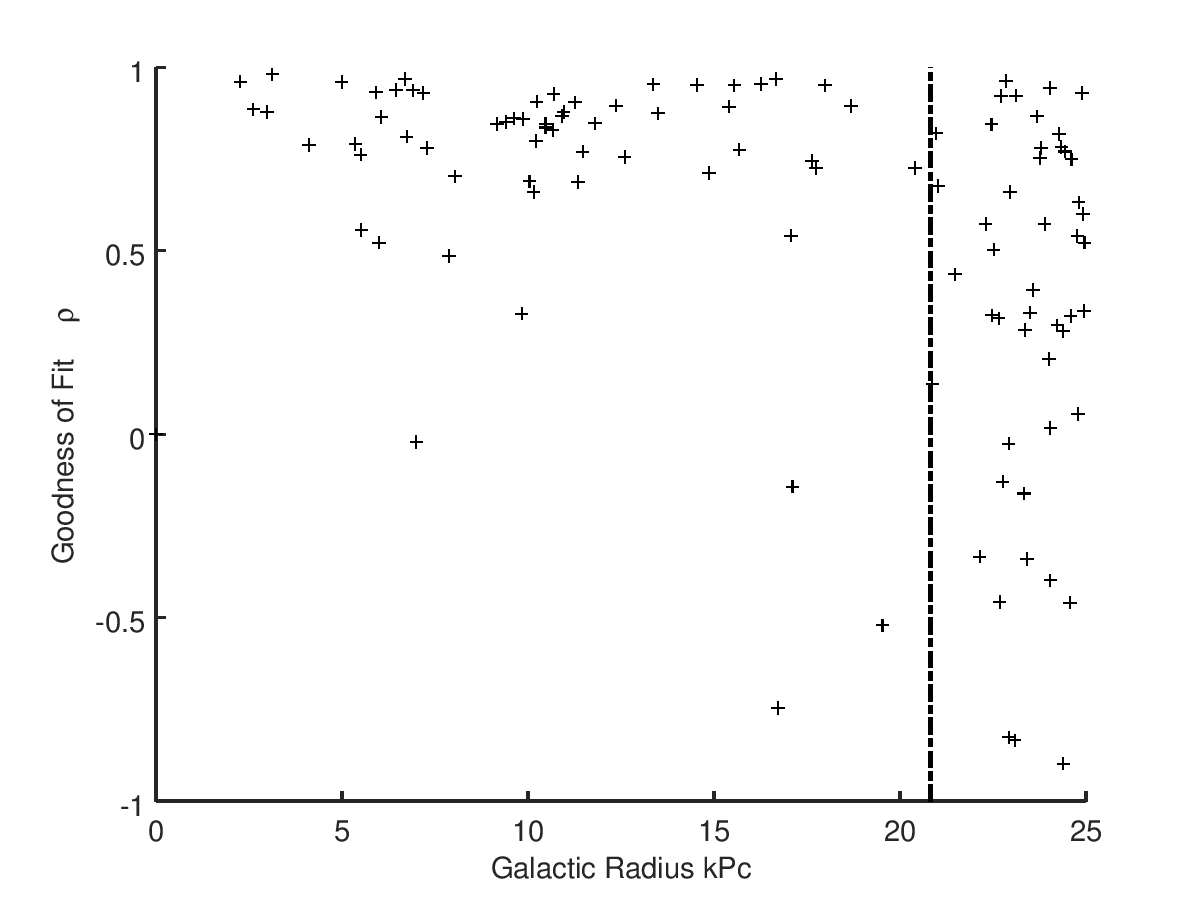}
		\caption{Correlation Coefficients versus Galaxy Radius}
		\label{fig:rho}
	\end{subfigure}
	~ 
	\begin{subfigure}[t]{0.45\textwidth}
		\centering
		\includegraphics[scale=0.4]{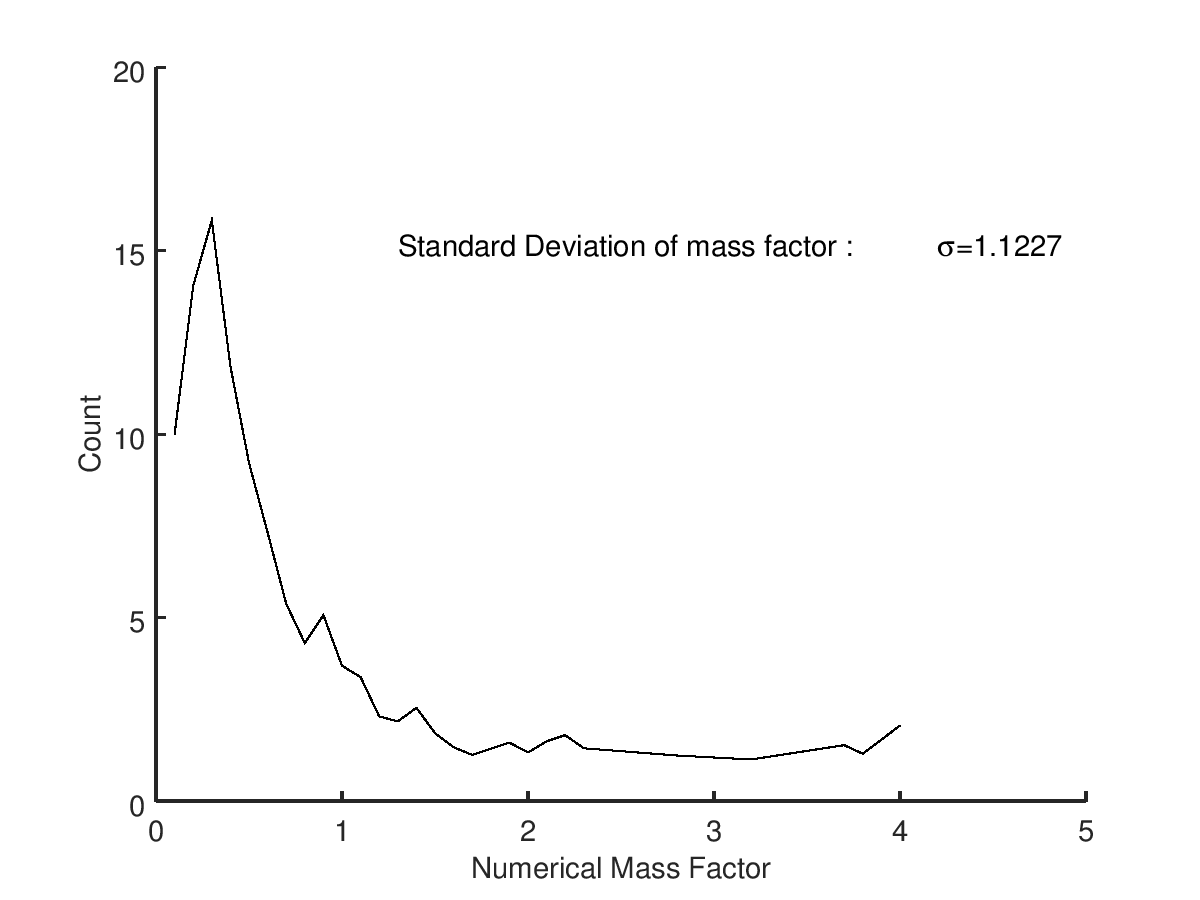}
		\caption{Distribution of Computed Mass Factor}
		\label{fig:MFD}
	\end{subfigure}
	~ 
	\begin{subfigure}[t]{0.45\textwidth}
		\centering
		\includegraphics[scale=0.4]{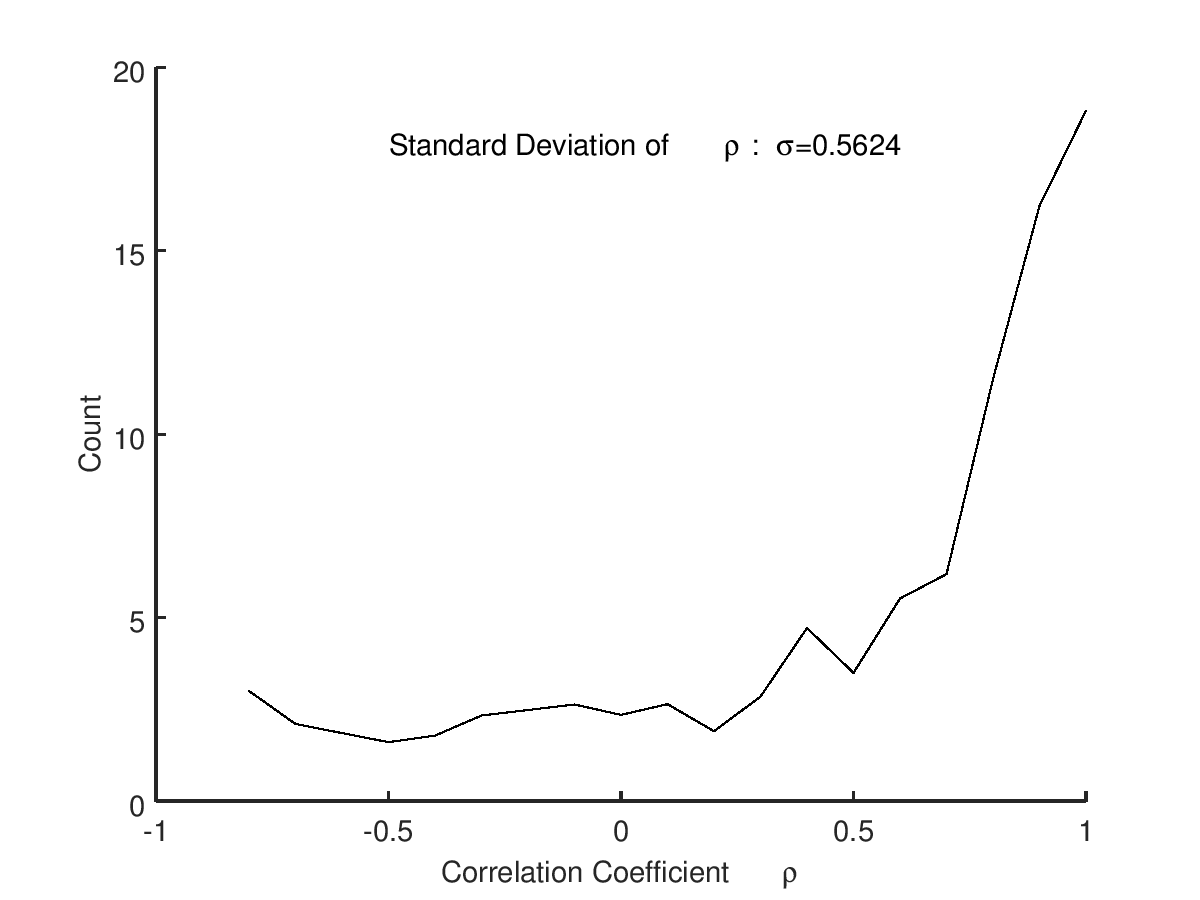}
		\caption{Distribution of Correlation Coefficient}
		\label{fig:rhoD}
	\end{subfigure}
	\caption{Analysis of variations of both mass factor and correlation coefficient across SPARC datasets.}
	\label{fig:distribs}
\end{figure*}

The QMD curve is more effective  at reproducing the form of the rotation curve, and in Figures \ref{fig:MF}, \ref{fig:rho}, \ref{fig:MFD} and \ref{fig:rhoD} we summarize the analysis across the whole dataset, having discarded those galaxies for which SPARC has fewer than $10$ points of  measured data for both surface luminosity and velocity of rotation, and discard galaxies with inclination angles of $<30^{\circ}$.
This leaves us with a sample of $109$ galaxies from the SPARC dataset.
This analysis reveals two important results.
Firstly the cluster of high values of $\rho$ is borne out in Figures  \ref{fig:rho} and \ref{fig:rhoD}.
This would tend to indicate that the QMD curve is a good match across a wide range of galaxies for the measured rotation curve, with the vast majority of galaxies having a value of $\rho > 0.8$.
Secondly, the fitting of the curves to the data involved numerically optimizing a multiplicative mass factor for the obtained mass density from the use of $\Upsilon$.
The need for this is justified by the unknown inputs into Equation \eqref{eqn:force_full}.
This `Mass Factor' appears to cluster around a value of around $0.4$, which appears to be close to universal across a large portion of the dataset, reducing the free parameters needed to predict reasonable rotation curves.
%
%

%
%

\section{The Continuum Limit and  the Einstein-Hilbert Action}
\label{sec:curved}

All of our analysis has been conducted upon a discrete spacetime, whereas the currently prevailing theories of both relativity and quantum mechanics are framed in the familiar mathematical language of continuous manifolds, differential forms and fibre bundles.
There is however precedent for considering how the geometry of discrete spacetimes behaves, and passes to a continuum limit.
This framework is the `Regge' calculus \cite{Regge1961,Lewis1983}, which provides a rigorous treatment of curvature in a simplicial mesh.
It is in principle possible to analyze the mesh of QMD using these tools to establish equivalent field equations to those of General Relativity, but this remains to be addressed rigorously in this program of work.

We can perhaps though take an intermediate step.
Our model presupposes the existence of holes or defects in the mesh to represent the presence of matter.
As such the dynamical processes on the mesh are intimately governed by the number of holes in the QMD mesh.
In the continuum limit, that is if we allow $L_p \rightarrow 0$, the simplicial geometry of this mesh must retain topological homeomorphism to the continuous manifold obtained.
In particular, the number of holes in the mesh must equate to the holes in the continuous spacetime in some fundamental way, otherwise a number of know topological invariants between the two spaces would differ.

Of course, this is well studied and understood in the language of homotopy, and the homotopy group of the continuum and discrete geometries must be the same.
For an arbitrary mesh, calculating this group is challenging, but there are other invariants of the space that if homotopy is preserved should also be preserved.

It is possible to frame such invariants in familiar constructs such as curvature two forms.
For example, one such construction is from the Gauss-Bonnet theorem \cite{Nash1983,Nakahara1990}, which states for a manifold $M$, one can construct the Euler number of the space as follows:

\begin{equation}
	\chi(M)=\int_M e(M) \mbox{, }
\end{equation}
where $e(M)$ is the Euler class of the manifold.

In essence our constraint of maintaining topological equivalence requires that any process on the manifold must result in no change to $\chi(M)$, or in other words its variation must be identically zero.
The Euler characteristic is only defined on even dimensional manifolds, and for the simple case of an orientable manifold with the normal structure of a Tangent Vector space $T_p M$ defined at every point $p \in M$, it can be written as:

\begin{equation*}
	e(M)=\frac{1}{32 \pi^2} \epsilon^{ijkl} \mathcal{R}_{ij} \wedge \mathcal{R}_{kl} \mbox{, }
\end{equation*}
where the $\mathcal{R}_{ij}$ are the curvature two-forms, and $\epsilon^{ijkl}$ is the totally asymmetric tensor in four indices.
In a vacuum, any permissible metric must not change the value of the Euler number.
So, we can write the homeomorphism constraint, as a metric variation condition on the vac and arrive at the following expression:

\begin{align}
	\delta_{g} \chi(M) &= 0 \\
	\delta_{g} \Bigg [ \frac{1}{32 \pi^2} \int_M \epsilon^{ijkl} \mathcal{R}_{ij} \wedge \mathcal{R}_{kl} \Bigg ] &= 0
\end{align}

It is well known that such a term can be added to the Einstein-Hilbert action, as in four dimensions it is a total derivative which results in a boundary term that does not contribute to the dynamics.
However, this form of gravitational theory (the so-called `Lovelock' gravity \cite{Lovelock1971}) does indeed produce non-trivial dynamics in the case of  dimensions $D>5$.
Further, it is also known that the produced Lagrangian is more similar to the low-energy limit of some string theories than the traditional Einstein-Hilbert action.
This variation condition is  suggestive, and similar at least in form to the familiar Einstein-Hilbert action.
More importantly, whereas the Lagrangian is intentionally chosen to produce the desired dynamical equations, here it is derived from  first principle considerations of how the mesh would transition to a continuum in the limit that $L_p \rightarrow 0$.
Given that the Einstein-Hilbert action is the starting point for the derivation of the field equations of GR, this is a potential connection.
It is beyond the scope of the current program to try and draw a more direct link between QMD and GR in the continuum limit, but this perhaps points the way to demonstrating a formal equivalence.

\section{Conclusion}
\label{sec:conclusion}

In this paper we have outlined a potential pathway from an emerged quantum geometry to both dynamics and a gravity like interaction.
This has relied upon the analysis of the behavior of a quantized translation operator, and the entropy of the underlying mesh as matter quanta move relative to each other.
The results have some interesting features, including the natural emergence of both Newtonian equations of motion and a gravity like attractive force between matter quanta.

We noted that the equations of gravitational attraction introduce a Yukawa correction with a natural length scale that operates on cosmic scales, along with a $1/r^3$ correction term, consistent with the field equations of GR.
As such the analysis motivated the use of the recent galactic velocity dispersion data from the SPARC collaboration, to test the effectiveness of the theory.
The results obtained reproduce many of the features of the experimental data without the recourse to the introduction of dark matter, or MOND like modifications to the operation of the gravitational force.
Although far from a perfect test of the equivalence of the QMD force law to the gravitational field from GR or Newtonian mechanics it is an encouraging result. 
Further, because of the scale factor, at low values of $r$ the modifications are small and we recover standard Newtonian gravity.

Beyond this analysis it is also possible to at least imagine the pathway from the mesh to the field equations of GR in the continuum limit.
Although this work does not cover this link in depth, a very top level analysis of how defect number conservation could lead to a variational condition on the curvature two forms of Riemannian geometry from which this link could be established.
Much remains to be done to develop the theory, including understanding more deeply this potential connection between the quantum mesh behavior and the low energy continuum limit. 
This and further experimental analysis will form the basis of future research.

\bibliographystyle{abbrv}
\bibliography{QuantumMeshGravity-arx}  

\end{document}